\documentclass[journal]{IEEEtran}
\usepackage{hyperref}
\usepackage{graphicx,graphics}
\usepackage{wasysym}
\usepackage{csquotes}
\usepackage{cases}
\usepackage{cite}

\begin{document}

\title{Improved spectral stability in spin transfer nano-oscillators: single vortex versus coupled vortices dynamics.}

\author{\IEEEauthorblockN{Nicolas Locatelli\IEEEauthorrefmark{1},
R. Lebrun\IEEEauthorrefmark{1}, V.V. Naletov\IEEEauthorrefmark{2}$^{,}$\IEEEauthorrefmark{3}$^{,}$\IEEEauthorrefmark{1}$^{,}$\IEEEauthorrefmark{4}, A. Hamadeh\IEEEauthorrefmark{2}, \\G. De Loubens\IEEEauthorrefmark{2}, O. Klein\IEEEauthorrefmark{2}$^{,}$\IEEEauthorrefmark{3}, J. Grollier\IEEEauthorrefmark{1}, and V. Cros\IEEEauthorrefmark{1}}

\IEEEauthorblockA{\IEEEauthorrefmark{1}Unit\'e Mixte de Physique CNRS/Thales, 1 av A. Fresnel, Campus de l'Ecole Polytechnique, \\91767 Palaiseau, France, and Univ Paris Sud, 91405 Orsay, France}

\IEEEauthorblockA{\IEEEauthorrefmark{2}Service de Physique de l'\'Etat Condens\'e (CNRS URA 2464), CEA Saclay, 91191 Gif-sur-Yvette, France}

\IEEEauthorblockA{\IEEEauthorrefmark{3}SPINTEC, UMR CEA/CNRS/UJF-Grenoble 1/Grenoble-INP, 38054 Grenoble, France}

\IEEEauthorblockA{\IEEEauthorrefmark{4}Institute of Physics, Kazan Federal University, Kazan 420008, Russian Federation}

\thanks{Present address of N. Locatelli is : Institut d'\'Electronique Fondamentale, Univ Paris-Sud, UMR CNRS 8622, Orsay, France. Corresponding authors: N. Locatelli (email: nicolas.locatelli@u-psud.fr) \& V. Cros (email: vincent.cros@thalesgroup.com).}}

\maketitle

\begin{abstract}
We perform a comparative study of spin transfer induced excitation of the gyrotropic motion of a vortex core with either uniform or vortex spin polarizers. The microwave output voltage associated with the vortex dynamics,  detected in both cases, displays a strong reduction of phase fluctuations in the case of the vortex polarizer, with a decrease of the peak linewidth by one order of magnitude down to $200$kHz at zero field. A thorough study of rf emission features for the different accessible vortex configurations shows that this improvement is related to the excitation of coupled vortex dynamics by spin transfer torques.
\end{abstract}


\IEEEpeerreviewmaketitle

\bstctlcite{IEEEexample:BSTcontrol}

\section{Introduction}

The spin transfer torque (STT) phenomenon~\cite{Berger_PRB54_1996, Slonczewski_JMMM159_1996}, that has been an intense research topic in the last decade, has allowed a crucial breakthrough for low energy writing in STT-MRAM~\cite{Khvalkovskiy_JPD46_2013}. Besides being an efficient tool for magnetization reversal, spin torque can also allow the generation of self-sustained magnetization oscillations in the GHz range. Thanks to the magneto-resistance (MR) effect, such magnetization dynamics is converted into an emitted voltage at the same frequency. This led to the proposal of new spintronics microwave devices called spin-transfer nano-oscillators (STNOs)~\cite{Kiselev_N425_2003}. The recent implementation of spin transfer dynamics in magnetic tunnel junctions (MTJ) has allowed to increase the output power of such oscillators close to the requirement for applications~\cite{Villard_IJSC45_2010, Dussaux_NC1_2010}. Concurrently, the introduction of \textit{vortex-based} spin transfer oscillators (STVOs) has allowed to significantly improve the spectral purity of the oscillation~\cite{Pribiag_NP3_2007, Locatelli_APL98_2011, Dussaux_NC1_2010, Hamadeh_PRL_2013, Grimaldi_PRB89_2014}. However further advances regarding these two issues are still required to eventually allow the use of STNOs in future telecommunication devices. Several applications are already considered today, among which their integration as nano-sized reference oscillators for FM signal modulation~\cite{Locatelli_NM13_2014, Manfrini_JAP109_2011, Iacocca_PRB85_2012, Pufall_APL86_2005, Muduli_PRB81_2010}, sensors for HDD read-heads~\cite{Suto_APE4_2011, Nagasawa_JAP111_2012}, or the design of interacting oscillators network in neuro-inspired processor architectures~\cite{Csaba_TIWCNNTAC_2012, Levitan_TIWCNNTAC_2012}.

A magnetic vortex is a magnetic state appearing in magnetic dots with a magnetization curling in the film plane and popping out-of-plane in a small region in its center called the \enquote{vortex core}, which size is comparable to the exchange length~ \cite{Feldtkeller_PKM1_1965}. A vortex is characterized by two parameters: its \textit{chirality} $C=\pm1$ giving the direction of the curling magnetization, and its \textit{core polarity} $P=\pm1$ giving the direction of the out-of plane core magnetization~\cite{Guslienko_JNN8_2008}. In this study, we are interested in the microwave signal generated by the spin-transfer induced excitation of the vortex \textit{gyrotropic mode}. This latter mode, the lowest in energy, corresponds to a gyration of the vortex core around its equilibrium position. Despite its small size, the vortex core has a crucial impact as it defines the frequency and direction of the gyrotropic oscillations. Because of the particular topology of the vortex ground state, one expects that the excitation of the gyrotropic mode by a dc current in a nano-pillar depends also on the magnetic configuration adopted by the polarizing layer. Several cases have been theoretically proposed to achieve self-sustained gyrotropic oscillations of the vortex core. The spin-current - and thus the polarizing layer - might be uniformly polarized, in the layer's plane~\cite{Sluka_JPDAP44_2011} or perpendicularly~\cite{Ivanov_PRL99_2007, Khvalkovskiy_PRB80_2009, Gaididei_IJQC110_2010}, or circularly-polarized, possibly with a vortex-like distribution~\cite{Khvalkovskiy_APL96_2010, Sluka_PRB86_2012}.

Our main objective is to compare, for given device dimensions, the output rf signal characteristics of vortex oscillations for two different magnetic configurations of the polarizing magnetic layer, either (i) in plane magnetized or (ii) vortex structure, that provides the spin polarization being the source of spin torque acting on the gyrating vortex. After discussing the accessible magnetic configurations, we show that the spectral quality of the output signal strongly depends on the actual magnetization distribution of the polarizer. Notably, we demonstrate that the spectral quality is improved by one order of magnitude when a second vortex is nucleated in the polarizing layer rather than uniform magnetization. Then, we explore the possibilities offered by vortex polarity switching. The critical influence of the relative polarities of the free and polarizer vortices is raised. This study demonstrates the importance of considering coupled mode in order to reduce noise-induced phase-fluctuations in STT-induced dynamics~\cite{Lebrun_PRA_2014}.

\section{Determination of the magnetic configuration by transport study}

\begin{figure}[t]
	\centering
	\includegraphics[width=\linewidth]{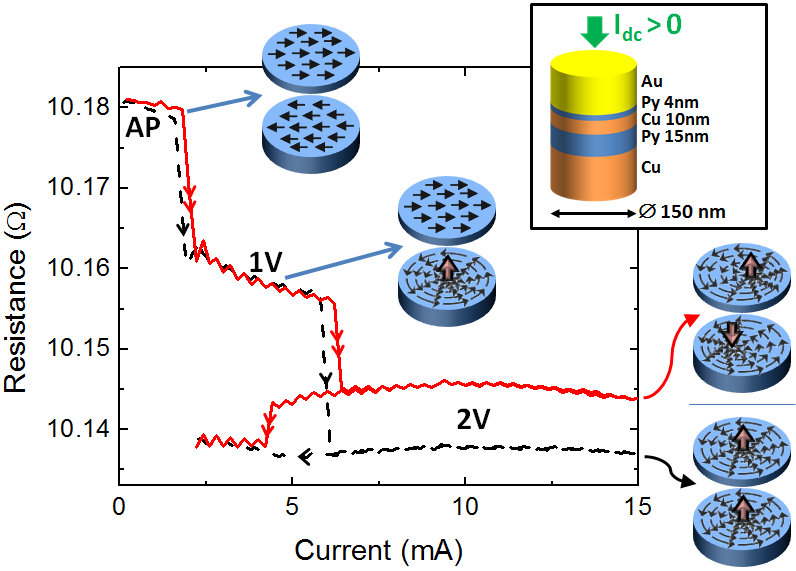}
	\caption{Resistance evolution versus dc-current for a positive current $I_{dc}>0$, measured at zero external magnetic. Schematics represent the magnetic configuration in thin Py(4nm) and thick Py(15nm) layers, big arrows correspond to vortices polarities. The two curves represent two possible path measured in repeated experiments. The red continuous line is associated with the observation of two-vortices state with opposite polarities, whereas the black dashed curve is associated with the observation of two-vortices state with identical polarities. For clarity, the increase of resistance associated with Joule heating was substracted. Insert : Sample geometry and positive current definition.}
	\label{fig:RvsIdc}
\end{figure}

The spin valve based STNO under study is composed of a Cu(60nm)/Py(15nm)/Cu(10nm)/Py(4nm)/Au(25nm) (Py=Ni$_{80}$Fe$_{20}$) stack patterned in a $\diameter150$nm circular nanopillar by standard e-beam and ion etching lithography process (see inset in Fig.~\ref{fig:RvsIdc}). The pillar dimensions and the thickness of the magnetic layers are chosen so that, in each layer, it will be possible to change the magnetic configuration between two stable states, namely in-plane uniform and vortex. The stability of the vortex configuration can be increased through the amplitude of the Oersted-field while the current flowing through the pillar is increased~\cite{Urazhdin_PRB73_2006}. We recall that the thicker the layer, the more stable the vortex state. In our convention, a positive current is flowing through the pillar, corresponding to electrons flowing from the thick Py(15nm) to the thin Py(4nm).

We first study the modification of the magnetic state in each layer by monitoring the nano-pillar resistance as the current is swept from $0$ to $15$mA and back to $2$mA. The averaged resistance versus dc current $I_{dc}$ measured at zero external magnetic is shown in figure~\ref{fig:RvsIdc}. At zero field and low $I_{dc}$, the stable state corresponds to anti-parallel uniform magnetizations, due to the dipolar interaction between the two magnetic layers in the nanopillar. As usual in GMR devices, such AP state is the higher resistance state. As the current is increased, a first resistance drop is observed at $I_{dc}=2$mA, approximately equal to half the full magnetoresistance ($\Delta R$), associated with a vortex nucleation in the thicker layer while the magnetization in the Py(4nm) layer remains quasi-uniform. This intermediate resistance state is then conserved up to about $I_{dc}=6$mA, when a second drop of resistance corresponding to the nucleation of a second vortex in the thin layer is observed. As they are both formed under the influence of the Oersted field, the chiralities of both vortices are identical and have an orthoradial symmetry similarly to the Oersted field : $C_{15nm}=C_{4nm}$. This is consistent with the lower resistance value observed associated with parallel magnetizations. Note that as the current is then swept back, both vortices remain stable even below their nucleation current. 

Repeating such R(I) measurement, we found as shown in Fig 1, that the resistance level associated with the 2-Vortices (2V) state can indeed take two different values. Because these measurements are performed at zero field, the core polarity during the nucleation process is random (or influenced only by local defects). Thus the core polarity of the vortex in each layer can be oriented either \enquote{up} or \enquote{down}. In the regime with a single vortex (1V), we see in Fig 1 that the resistance level is independent of the core orientation. However, we have to account for the possible configurations of the relative core polarities in order to understand that two resistance levels are observed in the range of current corresponding to the 2V state. Indeed, two distinct configurations can be obtained, corresponding to \textit{identical} (\textit{up-up} or \textit{down-down}) or \textit{opposite} (\textit{up-down} or \textit{down-up}) polarities.

To discriminate between these two configurations, we use the following procedure: we apply a large perpendicular field that favors identical polarities while injecting a large dc current of 10 mA that favors identical chiralities, then bring back the field to zero. In that case, we always observe the lower resistance state (black dashed curve in Fig.\ref{fig:RvsIdc}). Therefore, the lower resistance two-vortices state is associated with identical polarities, whereas the higher resistance state is associated with opposite polarities (red continuous curve). Our recent study of the evolution of the switching of core polarity with current, corroborating these results, is presented elsewhere~\cite{Locatelli_APL98_2011, Locatelli_APL102_2013}.

Note finally on Fig.\ref{fig:RvsIdc} that when we sweep back the current towards zero we find that the resistance corresponding to opposite polarities drops at $I_{dc}=4$mA and reaches the same level as the one for two vortices having identical polarities. We will see in next section that the resistance difference between these two 2V states is linked to differences in their dynamics.

\section{Characteristics of rf emission for the different magnetic configuration}

\begin{figure*}[t]
	\centering
	\includegraphics[width=.85\linewidth]{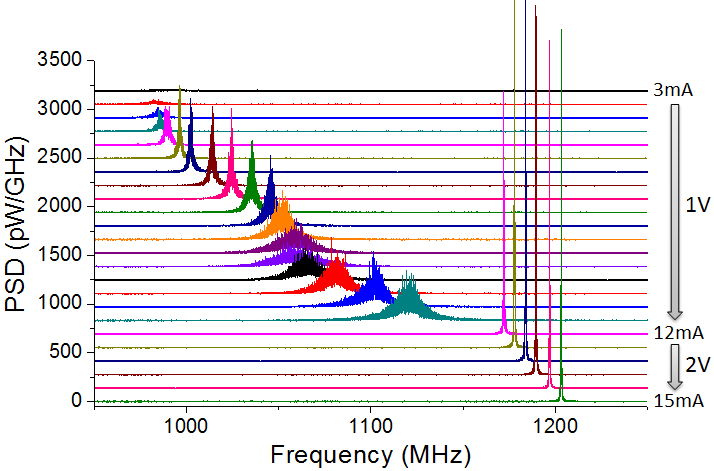}
	\caption{Power spectrum map measured for our spin valve STNO for increasing $I_{dc}$ in the case of a transition from a single vortex configuration to a two-vortices configuration with opposite polarities. The measurement is made under in-plane external field $B=34mT$. For this field value, the transition between the two configurations happens at $I_{dc}=11.5$mA. Peaks observed at harmonics of the fundamental frequency are not presented on this graph.}
	\label{fig:SpectraVsIdc}
\end{figure*}

\begin{figure}[!h]
	\centering
	\includegraphics[width=\linewidth]{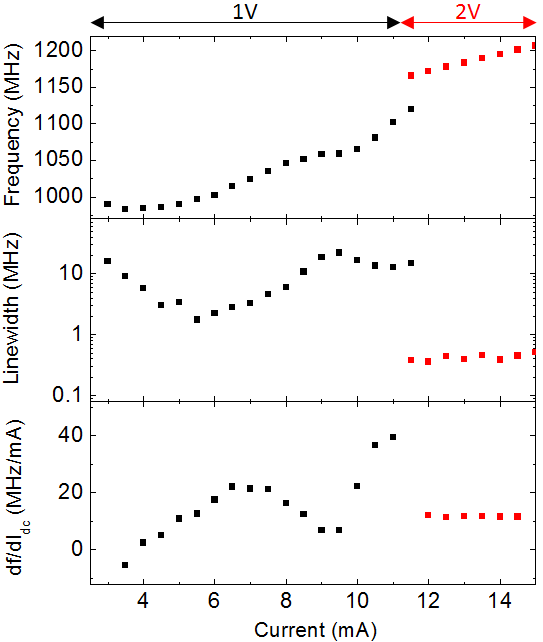}
	\caption{(a) Fundamental peak frequency, (b) linewidth and (c) tunability $df/dI_{dc}$ as a function of the injected dc current measured under B = 34 mT and at room temperature.}
	\label{fig:SpecsVsIdc}
\end{figure}

\subsection{In-plane polarizer versus vortex polarizer}

While sweeping $I_{dc}$, we also record the power spectrum of the nano-pillar voltage to compare the microwave spectra in the different configurations. For a positive current, it is expected to observe a STT-induced gyrotropic motion of the vortex in the Py($15$nm) layer, with the Py($4$nm) acting as the polarizing layer~\cite{Khvalkovskiy_APL96_2010}.

The power spectral density map, measured with a spectrum analyser after a $+30dB$ amplification is shown in Fig~ \ref{fig:SpectraVsIdc}. Here, an external in-plane field $B=34mT$ is applied to offset the 1V to 2V transition to higher current, ensuring to observe auto-oscillations under uniform polarizer well above the corresponding critical current. This field has shown no significant effect on the observed gyrotropic dynamics, as similar results could be obtained for smaller in-plane field values. We present the spectra measured while the injected current is increased, and focus and the transition from 1V to 2V with opposite polarities which occurs at 12 mA for the applied field value. We clearly observe two distinct regions on this map, which correspond to the 1V and 2V states. A striking modification of the microwave emission characteristics is then observed when the polarizing layer (thin layer) goes from in-plane magnetization to vortex magnetization.

In figure~\ref{fig:SpecsVsIdc}(a), we present the evolution of the signal frequency as a function of $I_{dc}$. The frequency is consistent with self-sustained gyrotropic oscillations of the vortex core in the thick Py(15nm) layer, with a frequency close to the evaluated gyrotropic eigen frequency ($\mu_{0}M_{S}=0.96$T and aspect ratio $\beta=L/R=0.2$ gives $f=\frac{5}{18\pi}\gamma\mu_{0}M_{S}\beta=952$MHz~\cite{Naletov_PRB84_2011,Guslienko_JAP91_2002}), augmented by Oersted-field confinement and spin-transfer action.
Figure~\ref{fig:SpecsVsIdc}(b) shows the corresponding evolution of the peak linewidth.
 
For $I_{dc}<11.5$mA, corresponding to the quasi-uniform polarizer condition, the frequency has a complex dependence versus $I_{dc}$. Moreover, we measure a correlation between between the $df/dI_{dc}$ (see Fig.~\ref{fig:SpecsVsIdc}(c)) and the linewidth dependences~\cite{Kim_PRL100_2008, Slavin_ITM45_2009, Grimaldi_PRB89_2014, Quinsat_APL97_2010, Muduli_PRL108_2012}. These observations are consistent with a significant contribution of the non-linear frequency shift to the increase of the linewidth in this configuration. Note that in this regime, the linewidth never goes below 1 MHz, similarly to the other results published in the literature for spin transfer vortex excitations~\cite{Pribiag_NP3_2007, Mistral_PRL100_2008, Dussaux_NC1_2010, Sluka_JPDAP44_2011}.

Upon the nucleation of the second vortex occurring around $I_{dc}=11.5$mA, we observe a sudden $50$MHz jump of the frequency and the frequency dependence versus $I_{dc}$ becomes highly linear. Concurrently, we observe a drastic decrease of the spectral linewidth, which now reaches values down to $200$kHz. The measured $df/dI_{dc}=13.3$MHz/mA linear increase of the frequency would then be dominated by the Oersted-field contribution to the vortex confinement~\cite{Khvalkovskiy_PRB80_2009}, and a strong decrease of the non-linear frequency shift under the new spin-transfer action could be consistent with the strong increase of spectral purity. 

It should be noticed that the linear dependence of the frequency with current in the 2V configuration is also a major asset for frequency modulation applications~\cite{Iacocca_PRB85_2012, Pufall_APL86_2005, Muduli_PRB81_2010}. These features are not modified by the in-plane external field application, and were also observed in the absence of any external field. Finally we emphasize that these measurements yield the lowest linewidth, $200$kHz, measured at zero field and room temperature so far for any type of STNOs.

\subsection{Two-vortices (2V): parallel vs opposite core polarities}

\begin{figure*}[t]
	\centering
	\includegraphics[width=\linewidth]{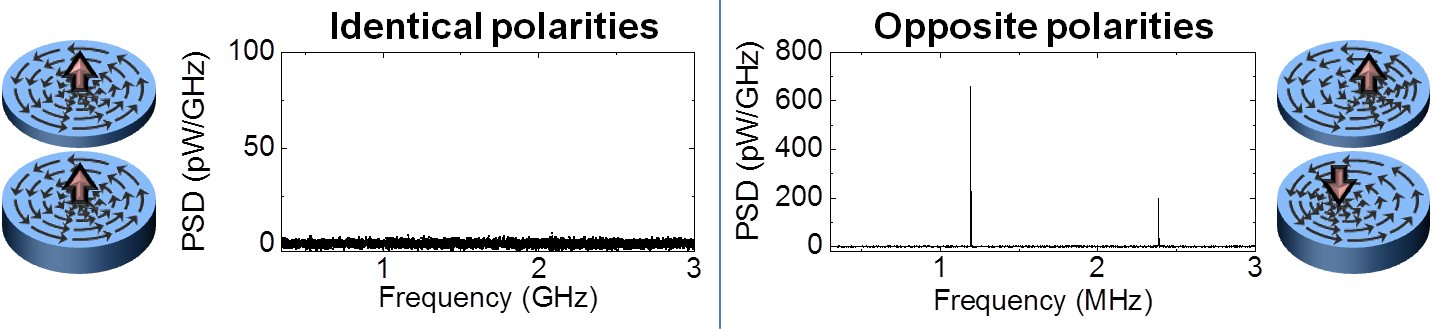}
	\caption{Spin-valve voltage spectrum measured in the two-vortices configuration in the case of (a) identical and (b) opposite polarities, both measured at $I_{dc}=15mA$ and zero magnetic field.}
	\label{fig:SpectraPvsAP}
\end{figure*}

As already mentioned, two different states (resistance levels) depending on the relative core polarities - identical or opposite polarities - of the two vortex configuration can be found at zero field. Interestingly, these two configurations yield very different microwave output signals. As demonstrated in figure~\ref{fig:SpectraPvsAP}, contrary to the case of opposite polarities we do not detect any rf signal for the parallel core configuration. 

This observation is consistent with the difference in average resistance measured between the two configurations shown in Fig.\ref{fig:RvsIdc}. Indeed, the low average resistance of the parallel core configuration is consistent with the absence of dynamics since in that case the two vortices are  perfectly centered and superimposed. On the contrary, the appearance of large amplitude vortex dynamics in the opposite core configuration increases the average resistance as the gyrotropic trajectory offsets the core from the center. Below the critical current $I_{dc}=4$mA, the oscillations in the opposite polarities configuration stop and the average resistance goes back to the lowest level. Between $I_{c}=4$mA and $15$mA, the resistance difference between the two configurations is constant, suggesting that the STT-induced amplitude of the gyrotropic oscillations does not change when the current amplitude increases.

The latter experimental results demonstrate the importance of the relative polarities of the polarizer vortex and free vortex for spin torque induced dynamics. Several arguments can be proposed to explain these observations. 

The main contribution to the STT action comes from the circular in-plane polarization~\cite{Khvalkovskiy_APL96_2010}. However, when the moving core is oscillating with low amplitude in a region close to the core of the vortex polarizer, the perpendicular contribution of the latter to the STT will be increasing or decreasing the total efficiency of the torque depending on its polarity~\cite{Sluka_PRB86_2012}. 
Second, the core-core dipolar interaction can also have a significant contribution when the two cores are close, changing sign depending on the respective core polarities. As illustrated in Fig.~\ref{fig:SpectraPvsAP}, adopting a configuration where the two cores are opposite produce a dipolar repulsion of the 2 cores, hereby breaking the axial symmetry of the system.
These two contributions have a significant influence on the determination of the critical current density necessary to start the gyrotropic oscillations, potentially explaining the strong differences between the two configurations. It should finally be noticed that the strong magnetization gradient associated with the vortex distribution in the core also strongly impacts the spin transport~\cite{Urazhdin_PRB73_2006} through the spin-valve, questioning the assumption of uniform spin polarization amplitude made in any theoretical study so far.

\subsection{Two-vortices (2V): Excitation of a coupled vortex mode}

Recent studies demonstrated that linewidth reduction can be obtained by STT-excitation of a coupled mode of the two magnetic layers~\cite{Gusakova_APL99_2011, Hamadeh_PRL_2013, Kudo_JAP111_2012, Lee_SR2_2012}, the polarizing layer being also involved in the STT-induced dynamics. Several features that we observed, such as the importance of relative core polarities, the frequency jump and the drastic linewidth reduction at the transition from 1 to 2 vortex configuration, suggests that the dynamics observed with two vortices of opposite core polarities corresponds to a spin transfer driven oscillations of a coupled gyrotropic motion of the two vortices~\cite{Sluka_PRB86_2012, Lebrun_PRA_2014}. 

Considering the existing dipolar coupling between the two vortices, their gyrotropic modes hybridize to produce two coupled eigen modes~\cite{Guslienko_APL86_2005}. Given the symmetry of the spin transfer action, one of these modes is self-sustained (here the mode dominated by the thick layer's vortex) while the second one (dominated by the thin layer's vortex) is over-damped~\cite{Sluka_PRB86_2012}. Under appropriate conditions energy transfer between the two modes can occur and help absorbing magnetic fluctuations around the self-sustained trajectory, artificially increasing the effective damping rate for amplitude fluctuations~ \cite{Lebrun_PRA_2014}. Thus, we note that whatever is the system geometry (diameter, thicknesses), relative vortices polarities then reveals to be a crucial parameter defining the symmetry of the excited mode.

Let us finally note that in the presented spin valve case the improved spectral quality of this coupled vortex mode is accessible only for opposite core polarities, thus at the cost of an efficient control of the magnetic configuration. However, we recently demonstrated that an easy and efficient control on the polarities can be done through the combined application of current and field in two vortices systems~\cite{Locatelli_APL102_2013}. The stated issue could then be turned into benefit by taking advantage of the possibility to switch on and off the oscillations, and the ac-voltage thereof, by switching one of the two polarities. This could be used as a calculation media in oscillator based transistor operations, or in oscillators based processors architectures, such as the recently proposed associative memories~\cite{Csaba_TIWCNNTAC_2012, Levitan_TIWCNNTAC_2012}.

\section{Conclusion}

In conclusion, we demonstrate the advantage of having the opportunity to tune a magnetic transition between single and double-vortex states in a spin-valve STNO to observe the evolution of the spectral quality obtained for both configurations. We show a one order of magnitude reduction of the peak linewidth. The evolution of frequency and linewidth with current are compatible with reduced non-linear frequency shift for the coupled vortex mode, concurrent with an increase of the effective damping rate for amplitude fluctuations. This results in a strongly improved stability, with an effective quality factor raising up to $f/\Delta f=6000$ without the need for external magnetic field. Interestingly, the rf characteristics associated with the 2V case present a frequency tunability that is almost linear and mainly associated with the change of the vortex confinement due to the Oersted field, with high potentiality for frequency control and modulation. Both features strengthen the potential applicability of vortex-based spin transfer oscillators in the near future.

\section*{Acknowledgments}
The authors acknowleges C. Deranlot, C. Ulysse and G. Faini for film growth and sample fabrication.
This research was partly funded by the French ANR (grant SPINNOVA ANR-11-NANO-0016) and the EU (FP7 grant MOSAIC ICT-FP7-317950). The work of V.V. Naletov is performed according to the Russian Government Program of Competitive Growth of Kazan Federal University.

\bibliographystyle{IEEEtran}
\bibliography{IEEEabrv,OneVsTwoVortex}

\end{document}